# Resource Letter TTSM-1: Teaching thermodynamics and statistical mechanics in introductory physics, chemistry, and biology

Benjamin W. Dreyfus[*] and Benjamin D. Geller[*]

*Department of Physics, University of Maryland, College Park, Maryland 20742*

David E. Meltzer

*Mary Lou Fulton Teachers College, Arizona State University, 7271 E. Sonoran Arroyo Mall, Mesa, Arizona 85212*

Vashti Sawtelle

*Department of Physics, University of Maryland, College Park, Maryland 20742*

## Abstract

This Resource Letter draws on discipline-based education research from physics, chemistry, and biology to collect literature on the teaching of thermodynamics and statistical mechanics in the three disciplines. While the overlap among the disciplinary literatures is limited at present, we hope this Resource Letter will spark more interdisciplinary interaction.

---
[*] These authors contributed equally to this work.

## I.    INTRODUCTION

Thermodynamics is central to our understanding of physics, chemistry, and biology. However, in most cases these three disciplines treat the topic in distinct ways. Each area has its own discipline-based education research literature, and rarely do we see researchers drawing upon work from disciplines other than their own. The primary goal of this Resource Letter is to provide a listing of peer-reviewed journal articles reporting research on teaching and learning of thermodynamics in physics, chemistry, biology, or any combination of those fields. A second goal is to delineate the differences and draw attention to places of potential overlap in the thermodynamics education research done in each of the three disciplines. We have focused on introductory-level thermodynamics (first-year physics, biology, and chemistry), but acknowledge that there is potential overlap at upper-level classes as well.

Instructors of introductory physics courses that include a discussion of thermodynamics are the primary intended audience of this Resource Letter. Our review of research on the teaching and learning of thermodynamics can help such instructors attend to concepts found to be especially challenging by introductory students. More specifically, it is increasingly common for introductory physics courses to be targeted at life science students, and interdisciplinary research, such as in biophysics, is of increasing importance to the physics community. Instructors teaching life science students or physics courses with biological applications should have familiarity with how thermodynamic ideas are discussed in biology and chemistry as well as in physics. The focus on thermodynamics learning in chemistry and biology should aid instructors' efforts to communicate effectively with majors in fields outside of physics, especially physics instructors explicitly engaged in developing interdisciplinary courses that overlap with chemistry and/or biology. Those who teach chemistry and biology courses that include discussions of thermodynamics would also find this paper helpful. Finally, this paper will serve as a primary reference for those who now or in the future carry out research on the teaching and learning of thermodynamics in any of these disciplines.

In choosing articles to include in this Resource Letter, we began with work done over 15 years by one of the authors on this paper (Meltzer). The starting bibliography (published at http://physicseducation.net/current/thermo_bibliography.pdf) was compiled through systematic searching of every research paper, book, conference proceeding, or dissertation related to research on the teaching of thermodynamics. To supplement this bibliography, we used the Education Resources Information Center (ERIC), an online digital library of education research, which includes research journals from the Biology Education, Physics Education, and Chemistry Education fields. In searching the database we used terms from thermodynamics such as *thermodynamics, kinetic theory, chemical bonding, entropy, enthalpy, Gibbs, diffusion, energy, heat, temperature,* and *osmosis*. In completing this search we found thousands of articles; we then applied various selection criteria, detailed below, to arrive at the listing presented in this paper. We did not constrain the search to any specific set of journals, so long as the paper was indexed in the ERIC database. This Resource Letter draws from articles in over 30 peer-reviewed journals from across the disciplines.

This Resource Letter does not try to compile a complete list of all the work that has been done in thermodynamics, nor does it provide a historical accounting of these works. Rather, we performed a literature search to select a set of papers that would indicate the depth of research in particular areas and draw attention to places where research could be further expanded. In searching for these broad terms in the education literature, we found many more papers than could reasonably be included in a single Resource Letter. To produce a coherent and reasonably sized overview, we limited our focus to introductory university-level thermodynamics. We thus excluded papers with a focus on elementary-school science and "pre-disciplinary" works, but have included some work from secondary-school analyses that align well with the work on thermodynamics at the university level. We draw attention to these differing populations when writing the descriptions for the individual references. As this Resource Letter is intended to be a collection of works for future researchers, we have kept to a minimum works from conference proceedings, and have excluded unpublished works.

Additionally, while we found several works that describe innovative methods for presenting thermodynamics in particular situations, if the analysis does not include either data on how students understood the concepts, or instructional implications, we excluded those works. Our decision on whether or not to include papers that contain instructional implications (but do not include student data) depended on several factors. Firstly, we acknowledge that our selection criteria were different for the different disciplines. In order to get representation of the literature from biology, in addition to chemistry and physics, we include a number of papers from biology education that rely more on anecdotal data than on data obtained through more systematic methodologies. We believe that the inclusion of such papers is important for understanding the current state of thermodynamic education research across the disciplines. A number of papers in the section on osmosis, for example, are centered around instructor reflections on how an activity was implemented, and those instructors' senses for how the activity might be used more broadly. More generally, in making decisions about whether to include papers that do not report on systematically obtained data, we chose to include those with particularly *broad* instructional implications. We chose not to include papers that focus on very narrow content topics, but rather those that have the potential to influence how topics like entropy or heat might be taught in a wide variety of contexts and representations.

Because our aim is not to provide a list of all work related to the teaching of thermodynamics, but rather to select a set of papers indicating the depth of research that exists in particular areas, we do not include the many textbooks and websites that articulate various approaches to thermodynamics education. The popular textbook *Thermal Physics* by Daniel Schroeder, for example, has been widely adopted and provides a particular perspective on how best to introduce ideas of statistical and thermal physics. Similarly, Chabay and Sherwood's *Matter & Interactions* text has been supported by peer-reviewed journal articles on the use of the text as part of an effective instructional strategy. It would be interesting for future research to compare the effectiveness of courses that order the curriculum in alignment to Schroeder's or Chabay and Sherwood's texts to those that choose other approaches, but at present such research does not exist and it is not the aim of this Resource Letter to present a list of textbook alternatives.

Likewise, while an increasing number of websites now contain links to papers addressing the teaching of thermodynamics, it is not our intent to provide a thorough account of such sites. Two websites in particular, entropysite.oxy.edu and

energyandentropy.com, are worth mentioning because they provide useful collections of literature that may be of interest to education researchers and instructors of thermodynamics at various levels. Many of the papers included on those websites can also be found in this Letter.

## II. REFLECTIONS ON THE DISCIPLINE-BASED LITERATURE

Not surprisingly, the disciplines of physics, biology, and chemistry have developed highly diverse approaches and emphases in their respective literatures on thermodynamics education. Most introductory chemistry courses, for example, neglect gas-phase reactions and the role of $pdV$ work in reactions that depend on the first law of thermodynamics, whereas physics courses would more likely include these considerations. Biology and chemistry students are also more likely than physics students to encounter an overwhelming emphasis on constant-pressure processes and a corresponding conflation of heat and enthalpy. In this section we summarize the primary areas of focus for each of the succeeding disciplinary sections, and draw attention to the places where work in these areas could be broadened. Additionally, we point out potential research directions for exploring the overlap in content knowledge among the disciplines. At the moment, the education literatures on thermodynamics in physics, chemistry, and biology have rarely drawn upon one another's work, so we see this overlap as a potential place of exploration for future research.

A theme that reappears throughout the discipline-based literature on both energy and entropy is the role that language – and in particular ontological metaphors – plays in student understanding of thermodynamics.  For example, a number of papers address the issue of whether to treat energy as a substance, and the impact that such a choice has on students' conceptual understanding. Likewise, the literature addresses the many metaphors used to talk about entropy, examining each one by exploring its use in toy problems and simplified examples. The question of whether the "disorder" metaphor is an appropriate choice, and in what contexts, is one that comes up a number of times in both the chemistry and physics education literature, with several authors suggesting alternatives.

## A. Physics

Most of the physics education literature discussing energy in the context of thermodynamics covers the first law of thermodynamics in a way that focuses on issues and applications relevant to engineering, such as gases and pistons. The literature on the basic concepts of heat and temperature, by contrast, goes back for more than four decades and considers a wide-range of issues, from primary school understanding to theoretical understandings of students' ideas. Other topics related to energy have also been developed more recently, though to date their connection to the chemistry and biology literatures remains limited.

The physics education literature on entropy primarily addresses student difficulties surrounding heat engines and the Carnot cycle, often adopting an approach to the second law that focuses on efficiency. However, there are a number of papers on the role played by language and conceptual metaphors in the construction of ideas about entropy. These papers attend to the benefits and limitations of various entropic metaphors in the context of a number of toy problems and simplified examples. The role of probability and statistics is given little coverage, and there is almost no discussion of how entropy relates to enthalpy and free energy.  The physics education literature has so far devoted little attention to diffusion and osmosis.

## B. Biology

The biology education literature on energy in thermodynamics is very limited, and what there is focuses on biochemistry. A small amount of the work in biology concerns energy in a general sense, but at present it falls significantly short of the breadth of coverage in the physics literature.

Very little biology education literature addresses entropy or the second law, either at the macro or micro scale.  At least one paper in biology education focuses on the role that student understanding of randomness plays in the development of a coherent conceptual model of the second law, but the absence of a more extensive biology education literature on that subject is striking given the central role that randomness plays in biological systems.  There is also very little about the relationship between entropy, enthalpy, and free energy, despite the biochemical importance of those relationships. Although the biology education literature does describe student understanding of diffusion and osmosis from a phenomenological perspective, there is almost no discussion of the physical mechanism underlying these concepts.

## C. Chemistry

The chemistry education literature on thermodynamics focuses on the areas of thermochemistry and energy in chemical reactions.  There are several articles that look similar to those in physics, using the familiar language of pistons and heat engines. These latter articles might originate primarily from a physical chemistry perspective.

Much of the chemistry education literature reflects an interest in how familiarity with entropy and the second law contributes to a more thorough understanding of chemical equilibria and spontaneity. To that end, the relationship between entropy, enthalpy, and Gibbs free energy is discussed extensively. These three constructs – entropy, enthalpy, and free energy – form the cornerstone of chemistry education literature on student understanding of the second law. In the chemistry education literature there is a focus on constant-pressure processes for which enthalpy is equivalent to heat, which has the potential to contribute to confusions for students who study thermodyanmics in chemistry and physics courses. There is very little mention in this literature set of diffusion and osmosis.

## III.    LITERATURE ORGANIZED BY TOPICAL AREA

In this section, we organize our references by content area within thermodynamics, dividing those that primarily address energy from those that primarily address entropy and statistical mechanics. Further, as we reviewed the literature we found several seemingly related subareas of thermodynamics that did not commonly reference one another. We present the literature in a format that mirrors these distinctions. For example, while the topical area "Heat and Temperature" is conceptually related to an area like "The First Law of Thermodynamics," we find that these literatures do not typically make any explicit reference to one another, so we assign each of these sets of literature its own subsection. This is not true of all the literature, and we have also included a section entitled "Thermodynamics: General" to distinguish the literature set that discusses concepts that span multiple topics within thermodynamics (e.g., papers that discuss both the first and second laws). At times an article was written to primarily address one issue in thermodynamics, but also speaks to a secondary issue in the text. In these cases we have listed the article under its primary conceptual area, and included them in a "See Also" list for their secondary area.

As part of the goal of this Resource Letter is to examine which topics have been explored in the various disciplines, each section is further subdivided into disciplinary areas. In choosing to label individual works with a disciplinary heading we have attended to both the journal the article was published in and the audience the article was intended to reach. We recognize that in many cases the authors would not have labeled their work as relating to thermodynamics or statistical mechanics. Nonetheless, as our goal with this Resource Letter is in part to point to potential places of overlap between physics and other disciplinary areas, we have made judgments as to where these overlaps occur (e.g., biologists might not label osmosis as relating to statistical mechanics, but we have included work on osmosis here).

We also want to draw attention across disciplinary boundaries to the need for work that contains data about how students respond to and think about ideas in thermodynamics. As such, we have created a code that denotes an article that included student data **(SD)** in the article. We hope in this way to point to the need for additional research uncovering students' ideas in understanding thermodynamics within the different disciplinary contexts.

## A. Thermodynamics: General

These papers are concerned with thermodynamics as a whole, and are not classified into any of the more specific subtopics.  In physics, chemistry, and biology, people have thought about how thermodynamics relates to the rest of the introductory curriculum.

This includes several proposals to expand the role of thermodynamics in the introductory course and to integrate it with mechanics. We also include empirical studies on student reasoning in physics and engineering that cut across multiple thermodynamics topics, which include the development of concept inventories.

### 1. Biology

1. "Building a foundation for bioenergetics," E. Hamori, Biochemistry and Molecular Biology Education **30**, 296–302 (2002). Summarizes principles of thermodynamics, emphasizing a macroscopic and axiomatic approach, and applies them to bioenergetics. The principles of equilibrium thermodynamics are extended to nonequilibrium steady-state systems. (E)

### 2. Chemistry

2. "The teaching of thermodynamics at preuniversity level," I.F. Roberts and D.S. Watts, Phys. Educ. **11**, 277–284 (1976). Compares the thermodynamics topics covered in various secondary chemistry and physics curricula. (E)

3. "Difficulties of Students from the Faculty of Science with Regard to Understanding the Concepts of Chemical Thermodynamics," H. Sokrat, S. Tamani, M. Moutaabbid, and M. Radid, Procedia - Soc. Behav. Sci. **116**, 368–372 (2014). **(SD)** Identifies student difficulties with thermodynamics and possible reasons for these difficulties by administering a survey to college chemistry students in Morocco. Includes a discussion of linguistic and social dimensions of student difficulty. (E)

4. "Gathering Evidence for Validity during the Design, Development, and Qualitative Evaluation of Thermochemistry Concept Inventory Items," D. Wren and J. Barbera, J. Chem. Educ. **90**, 1590–1601 (2013). **(SD)** Looks at the design, development, and qualitative evaluation of concept inventory items for the Thermochemistry Concept Inventory (TCI). Evidence of content and response process validity are used to support the instrument's validity. (E)

5. "Psychometric analysis of the thermochemistry concept inventory," D. Wren and J. Barbera, Chem. Educ. Res. Pract. (2014), DOI: 10.1039/c3rp00170a . **(SD)** An analysis of the Thermochemistry Concept Inventory (TCI), intended to assess conceptual understanding of thermochemistry in general chemistry courses. (I)

6. "A review of research on the teaching and learning of thermodynamics at the university level," K. Bain, A. Moon, M.R. Mack, and M.H. Towns, Chem. Educ. Res. Pract. (2014). DOI: 10.1039/c4rp00011k . A review article synthesizing discipline-based education research on thermodynamics. (E)

See also: Ref. 122, Ref. 125

## 3. Engineering

7. "Preliminary results from the development of a concept inventory in thermal and transport science," B.M. Olds, R.A. Streveler, R.L. Miller, and M.A. Nelson, in *(CD) Proceedings, 2004 American Society for Engineering Education Conference* (2004). **(SD)** Results from testing the first version of the Thermal and Transport Concept Inventory (TTCI) for reliability and validity. (E)

8. "Concept inventories meet cognitive psychology: Using beta testing as a mechanism for identifying engineering student misconceptions," R.L. Miller, R.A. Streveler, M.A. Nelson, M.R. Geist, and B.M. Olds, in *Proceedings of the American Society for Engineering Education Annual Conference* (2005), pp. 12–15. **(SD)** Beta testing for the TTCI (see Ref. 7) demonstrates robust misconceptions about energy versus temperature and about equilibrium versus steady state. (I)

9. "Probing student understanding of basic concepts and principles in introductory engineering thermodynamics," C.H. Kautz and G. Schmitz, in *ASME 2007 International Mechanical Engineering Congress and Exposition* (2007), pp. 473–480. **(SD)** Clicker-question data from an engineering thermodynamics course reveal a number of conceptual difficulties related to work, heat, the first law, entropy, state vs. process quantities, and intensive vs. extensive quantities. (E)

10. "Identifying and repairing student misconceptions in thermal and transport science: Concept inventories and schema training studies," R.L. Miller, R.A. Streveler, D. Yang, and A.I. Santiago Román, Chem. Eng. Educ. **45**, 203–210 (2011). **(SD)** The misconceptions identified in the TTCI (see Ref. 7) are associated with ontological difficulties around emergent processes. A method for "schema training" was created to help students develop a mental model for emergence, and the data showed gains in student learning. (A)

### 4. Physics

11. "Bringing atoms into first-year physics," R.W. Chabay and B.A. Sherwood, Am. J. Phys. **67**, 1045–1050 (1999). Argues for building the introductory physics course around the atomic model of matter, rather than treating thermal physics as a separate topic from mechanics, and presents the approach used in the *Matter and Interactions* course. (E)

12. "Thermal physics in the introductory physics course: Why and how to teach it from a unified atomic perspective," F. Reif, Am. J. Phys. **67**, 1051–1062 (1999). Like Ref. 11, this paper also advocates teaching thermal physics from an atomic perspective in the introductory physics course, but presents thermal physics as a self-contained unit. A number of macroscopic thermodynamic results are derived from microscopic models, at a level appropriate for first-year physics students. (E)

13. "Teaching the photon gas in introductory physics," H.S. Leff, Am. J. Phys. **70**, 792–797 (2002). This paper discusses how one might supplement the traditional thermodynamic treatment of the ideal gas with an introductory-level treatment of the photon gas. An advantage is that ideas about modern physics can be incorporated into the thermodynamic discussion. (E)

14. "Investigation of student learning in thermodynamics and implications for instruction in chemistry and engineering," D.E. Meltzer, *AIP Conf. Proc.* **883,** 38–41 (2007). **(SD)** Students in first-year physics, upper-level thermal physics, and physical chemistry had difficulties on questions about work, heat, and entropy. Interdisciplinary implications are discussed for physics, chemistry, and engineering. (E)

See also**:** Ref. 136, Ref. 137, Ref. 166

### 5. Multidisciplinary

15. "Forms of Productive Complexity as Criteria for Educational Reconstruction: The Design of a Teaching Proposal on Thermodynamics," O. Levrini, P. Fantini, B. Pecori, and G. Tasquier, Procedia - Soc. Behav. Sci. **116**, 1483–1490 (2014). Discusses thermodynamic teaching materials designed for and implemented with secondary students in Italy. The literature on student difficulties around thermodynamics is reviewed, and the potential of the proposed curriculum to promote both intellectual and emotional growth is described. (E)

## B. Gas Laws and Kinetic Theory

These papers deal with macroscopic models of gases (gas laws), microscopic models (kinetic theory), and in some cases, the connections between the two. Though research in this topic is still limited, this is an area where the macro-micro connection (the connection between thermodynamics and statistical physics) can be explicated at the introductory level.

### 1. Chemistry

16. "The assessment of students and teachers' understanding of gas laws," H.-S. Lin, H.-J. Cheng, and F. Lawrenz, J. Chem. Educ. **77**, 235–238 (2000). **(SD)** Both teachers and advanced students displayed common misconceptions, including misuse of gas equations and failure to distinguish between system and surroundings. (E)

17. "The role of multiple representations in the understanding of ideal gas problems," S.P. Madden, L.L. Jones, and J. Rahm, Chem Educ Res Pr. **12**, 283–293 (2011). **(SD)** Introductory chemistry students tended to use a single representation to explain gas law phenomena, while showing difficulty with other representations. Upper-level students demonstrated more representational flexibility. (E)

18. "Pushing for particulate level models of adiabatic and isothermal processes in upper-level chemistry courses: a qualitative study," G.E. Hernández, B.A. Criswell, N.J. Kirk, D.G. Sauder, and G.T. Rushton, Chem. Educ. Res. Pract. (2014), DOI: 10.1039/c4rp00008k . **(SD)** Video analysis of class discussions showed that upper-level students inappropriately applied macroscopic gas laws in reasoning about thermodynamic processes, and struggled to use particulate models. (E)

### 2. Physics

19. "Students' reasonings in thermodynamics," S. Rozier and L. Viennot, Int. J. Sci. Educ. **13**, 159 – 170 (1991). **(SD)** Students oversimplify multivariable relationships, such as $pV = nRT$, by incorrectly reducing the number of variables and by introducing chronology and causality. (E)

20. "Designing a learning sequence about a pre-quantitative kinetic model of gases: the parts played by questions and by a computer-simulation," M. Méheut, Int. J. Sci. Educ. **19**, 647–660 (1997). **(SD)** A learning sequence is assessed on how effectively students adopt a particle model to explain properties of gases. (E)

21. "Teaching thermodynamics with Physlets® in introductory physics," A.J. Cox, M. Belloni, M. Dancy, and W. Christian, Phys. Educ. **38**, 433–440 (2003). The development of interactive computer simulations to connect macroscopic gas properties to microscopic particle models. (E)

22. "Student understanding of the ideal gas law, Part I: A macroscopic perspective," C.H. Kautz, P.R.L. Heron, M.E. Loverude, and L.C. McDermott, Am. J. Phys. **73**, 1055–1063 (2005). **(SD)** Student difficulties related to pressure, volume, temperature, and their relationship are documented, and the development of tutorials to address these difficulties is described. (E)

23. "Student understanding of the ideal gas law, Part II: A microscopic perspective," C.H. Kautz, P.R.L. Heron, P.S. Shaffer, and L.C. McDermott, Am. J. Phys. **73**, 1064–1071 (2005). **(SD)** Students displayed serious errors in the microscopic interpretation of the variables in the ideal gas law. (E)

See also**:** Ref. 111, Ref. 146.

### 3. Multidisciplinary

24. "An interdisciplinary study of student ability to connect particulate and macroscopic representations of a gas," K. Monteyne, B.L. Gonzalez, and M.E. Loverude, *AIP Conf. Proc.* **1064**, 163–166 (2008). **(SD)** Students in general-education physics and chemistry courses were more successful in going from the particulate to the macroscopic realm than vice versa. (E)

See also**:** Ref. 161

## C. Energy: General

In the science education and physics education literatures, there is an extensive body of work on the nature of energy and the learning and teaching of energy concepts. We have not attempted to include this entire literature here, but have narrowed our focus to energy as it relates to thermodynamics in and across the disciplines. We therefore have mostly excluded papers that are 1) focused exclusively on mechanical energy, without a thermodynamics connection, or 2) situated in K-8 science education, and therefore "pre-disciplinary." While physics predominates even within this narrower slice, biology and chemistry are also represented. There is an ongoing conversation, both theoretical and empirical, about the fundamental elements of the energy concept.

### 1. Biology

25. "Understanding of energy in biology and vitalistic conceptions," J. Barak, M. Gorodetsky, and D. Chipman, Int. J. Sci. Educ. **19**, 21–30 (1997). **(SD)** Students' understanding of energy in biology was significantly correlated with scientific as opposed to vitalistic (i.e., that biological phenomena cannot be explained by physics and chemistry) explanations. (E)

26. "Diagnosing students' understanding of energy and its related concepts in biological context," V.M. Chabalengula, M. Sanders, and F. Mumba, International Journal of Science and Mathematics Education **10**, 241–266 (2011). **(SD)** Biology students displayed conceptual difficulties when applying energy concepts in biological contexts. (E)

### 2. Chemistry

27. "Heat and work are not 'forms of energy,'" G.D. Peckham and I.J. McNaught, J. Chem. Educ. **70**, 103–104 (1993). Responds to problematic statements in textbooks, and argues that heat and work are processes, not forms of energy. (E)

### 3. Physics

28. "Some alternative views of energy," D.M. Watts, Phys. Educ. **18**, 213–217 (1983). **(SD)** Seven alternative frameworks for energy are identified from student interviews: human-centered, depository, ingredient, obvious activity, product, functional, and flow-transfer. (E)

29. "The concept of energy without heat or work," H.R. Kemp, Phys. Educ. **19**, 234–240 (1984). Advocates for a sequence of definitions in teaching energy (kinetic, potential, total, and internal energy), avoiding the concepts of heat and work. (E)

37. "Teaching Energy Conservation as a Unifying Principle in Physics," J. Solbes, J. Guisasola, and F. Tarín, J. Sci. Educ. Technol. **18**, 265–274 (2009). **(SD)** A teaching sequence for energy was developed that emphasizes energy conservation throughout all of physics (rather than just mechanics), and students were successful in applying energy concepts to a variety of situations. (E)

38. "Energy as a substancelike quantity that flows: Theoretical considerations and pedagogical consequences," E. Brewe, Physical Review Special Topics - Physics Education Research **7**, 020106 (2011). **(SD)** Presents a curricular framework for a Modeling Instruction course that uses an energy-as-a-substance conceptual metaphor, and examines episodes showing student use of energy conceptual resources. (I)

39. "Representing energy. I. Representing a substance ontology for energy," R.E. Scherr, H.G. Close, S.B. McKagan, and S. Vokos, Physical Review Special Topics - Physics Education Research **8**, 020114 (2012). **(SD)** The authors argue that using the substance ontology for energy is valuable in instruction because it supports energy conservation, transfer, and flow. (E)

See also: Ref. 153, Ref. 168

### *4. Multidisciplinary*

40. "Using metaphor theory to examine conceptions of energy in biology, chemistry, and physics," R. Lancor, Sci. & Educ. **23**, 1245–1267 (2014). Six substance metaphors for energy are identified in biology, chemistry, and physics textbooks: energy can be accounted for, can change forms, can flow, can be carried, can be lost, and can be stored, added, or produced. (I)

## D. Heat and temperature

The literature on heat and temperature spans the full range from studies of young children's understanding of "hot" and "cold" to university-level thermodynamics. Again, we have focused here on work that informs thermodynamics education in and across the disciplines, and have therefore left out much of the "pre-disciplinary" literature. The disciplinary literature in this area is extensive, and may be the most developed of any of the topics we include in this Resource Letter. There is a longstanding thread on the use of the word "heat": the distinction between heat and thermal energy, whether "heat" is a noun or a verb, and whether the terminology really matters for students' conceptual understanding.

## 1. Chemistry

41. "The definition of heat," T.B. Tripp, J. Chem. Educ. **53**, 782–784 (1976). Criticizes chemistry textbook language about heat, which confuses heat with kinetic energy or temperature, or says that systems "have" heat. The article operationalizes the definition of heat by reference to calorimeters. (E)

42. "Further reflections on heat," F.M. Hornack, J. Chem. Educ. **61**, 869–874 (1984). Four common errors found in science texts are identified: a system contains heat, changes associated with work are attributed to heat, a temperature change in an isolated system is attributed to heat, and heat is microscopic thermal energy. Notational confusion around heat as an inexact differential is discussed. (E)

43. "Understanding of elementary concepts in heat and temperature among college students and K-12 teachers," P.G. Jasien and G.E. Oberem, J. Chem. Educ. **79**, 889–895 (2002). **(SD)** There was no correlation between the number of semesters of college physical science and basic understanding of thermal equilibrium. (E)

44. "Student learning of thermochemical concepts in the context of solution calorimetry," T.J. Greenbowe and D.E. Meltzer, Int. J. Sci. Educ. **25**, 779–800 (2003). **(SD)** Exam and interview data on calorimetry problems showed difficulties in using the correct mass (system versus surroundings) and the correct sign for heat of reaction. (E)

45. "Can the study of thermochemistry facilitate students' differentiation between heat energy and temperature?" M. Niaz, Journal of Science Education and Technology **15**, 269–276 (2006). **(SD)** Results on the Test of Heat Energy and Temperature (THT) showed that students continued to have difficulties in differentiating heat energy and temperature after studying thermochemistry in an undergraduate chemistry course. (E)

See also**:** Ref. 125

## 2. Engineering

46. "Misconceptions about rate processes: Preliminary evidence for the importance of emergent conceptual schemas in thermal and transport sciences," R. Miller, M. Chi, M. Nelson, and M. Geist, in *ASEE Conference Proceedings* (2006). **(SD)** Results from the TTCI (see Ref. 7) show robust confusion between the rate and the amount of heat transfer. (E)

See also: Ref. 8

## 3. Physics

47. "The use and misuse of the word 'heat' in physics teaching," M.W. Zemansky, Phys. Teach. **8**, 296–300 (1970). As part of a decades-long conversation in the physics education literature, this article criticizes textbooks that refer to the "heat in a body" or use "heat" as a verb, and argues for introducing internal energy and the rigorous form of the first law. (E)

48. "The teaching of the concept of heat," J.W. Warren, Phys. Educ. **7**, 41–44 (1972). **(SD)** Both textbooks and students confuse heat with internal energy, and heat with molecular kinetic energy. (E)

49. "Teaching heat—an analysis of misconceptions," M.K. Summers, School Science Review **64**, 670–676 (1983). This article looks at "misconceptions" in physics textbooks. In contrast to Ref. 39, it argues that "heat" should not be used as a noun, but only "heating" as a verb to describe a process. (E)

50. "When heat and temperature were one," M. Wiser and S. Carey, in *Mental Models*, edited by D. Gentner and A. Stevens (Erlbaum, Hillsdale NJ, 1983), pp. 267–297. A historical account of the 17th-century "source-recipient model," in which heat and cold were separate concepts while heat and temperature were undifferentiated, and suggests that this model is parallel to present-day novice theories. (E)

51. "Secondary students' conceptions of the conduction of heat: Bringing together scientific and personal views," E.E. Clough and R. Driver, Phys. Educ. **20**, 176–182 (1985). **(SD)** Ideas about heat associated with young children's thinking persist into secondary school, including treating cold as a substance. (E) chapter presents a study addressing these conceptual issues with computer models linking the molecular and macroscopic levels. (E)

66. "'Is heat hot?' Inducing conceptual change by integrating everyday and scientific perspectives on thermal phenomena," M. Wiser and T. Amin, Learning and Instruction **11**, 331–355 (2001). **(SD)** The focus is on two ontologies for heat: heat as energy (the scientific usage) and heat as hotness (the everyday usage). In the intervention study, this metaconceptual issue was addressed explicitly. (I)

67. "Introductory thermal concept evaluation: Assessing students' understanding," S. Yeo and M. Zadnik, Phys. Teach. **39**, 496–504 (2001). **(SD)** Includes the full set of questions in the Thermal Concept Evaluation (TCE), and discusses development and validation. (E)

68. "Concerning scientific discourse about heat," D. Brookes, G. Horton, A. Van Heuvelen, and E. Etkina, *AIP Conf. Proc.* **790**, 149–152 (2005). **(SD)** Using grammatical analysis to identify ontologies, definitions of heat in physics textbooks are classified into substance- and process-based definitions. While physicists formally define heat as a process, they primarily use substance language. (I)

69. "Use of the thermal concept evaluation to focus instruction," G.R. Luera, C.A. Otto, and P.W. Zitzewitz, Phys. Teach. **44**, 162–166 (2006). **(SD)** The TCE (Ref. 67) was used as a diagnostic pretest to target specific areas of inquiry in a course for pre-service elementary teachers. Improvements are proposed to the TCE, to identify specific misconceptions. (E)

70. "Understanding the role of measurements in creating physical quantities: A case study of learning to quantify temperature in physics teacher education," T. Mäntylä and I.T. Koponen, Science & Education **16**, 291–311 (2006). **(SD)** The development of the concept of temperature was broken down into three stages: the level of qualities, the level of quantities and laws, and the level of structured theory. These levels were used to analyze students' network representations. (I)

71. "Students' pre-knowledge as a guideline in the teaching of introductory thermal physics at university," R. Leinonen, E. Räsänen, M. Asikainen, and P.E. Hirvonen, European Journal of Physics **30**, 593–604 (2009). **(SD)** A survey given to entering students in a thermal physics course showed difficulties with temperature and heat. In an adiabatic compression task, students use the ideal gas law rather than the first law. (E)

See also: Ref. 138

### 4. Multidisciplinary

72. "The concepts of heat and temperature: The problem of determining the content for the construction of an historical case study which is sensitive to nature of science issues and teaching–learning issues," K.C. de Berg, Science & Education **17**, 75–114 (2006). Presents the historical development of the concepts of heat and temperature, and promotes this as a valuable case study for teaching the nature of science. (E)

73. "Heat energy and temperature concepts of adolescents, adults, and experts: Implications for curricular improvements," E.L. Lewis and M.C. Linn, J. Res. Sci. Teach. **31**, 657–677 (1994). **(SD)** Interviews with students and nonscientist adults showed beliefs that metals and wool have active thermal properties. Scientists had difficulty explaining the difference between heat and temperature. (E)

### 5. Non-discipline-specific

74. "Heat and temperature," G. Erickson and A. Tiberghien, in *Children's Ideas in Science*, edited by R. Driver, E. Guesne, and A. Tiberghien (Open University Press, Milton Keynes, 1985), pp. 52–84. **(SD)** Results are synthesized from a number of studies on children's ideas about heat and temperature, and the development of those ideas with teaching. (E)

75. "On the thermal properties of materials: common-sense knowledge of Italian students and teachers," M.R. Sciarretta, R. Stilli, and M. Vicentini Missoni, Int. J. Sci. Educ. **12**, 369–379 (1990). **(SD)** Students' explanations of thermal phenomena differ from teachers' explanations, but this is because school science deals with equilibrium thermodynamics; students' "common-sense" intuitions were appropriate for entropy and irreversible processes. (E)

76. "Children's and lay adults' views about thermal equilibrium," M. Arnold and R. Millar, Int. J. Sci. Educ. **16**, 405–419 (1994). **(SD)** Students and adults had difficulty reasoning about a candle heating a tin of water, using an on/off model for heat. (E)

77. "A review of selected literature on students' misconceptions of heat and temperature," M. Sözbilir, Boğaziçi University Journal of Education **20**, 25–40 (2003). This review collects misconceptions about heat and temperature found in the science education literature. (E)

78. "Helping students revise disruptive experientially supported ideas about thermodynamics: Computer visualizations and tactile models," D. Clark and D. Jorde, J. Res. Sci. Teach. **41**, 1–23 (2004). **(SD)** A tactile model (visualizing how hot something feels) supported conceptual gains on thermal equilibrium. (E)

79. "Longitudinal conceptual change in students' understanding of thermal equilibrium: An examination of the process of conceptual restructuring," D.B. Clark, Cognition and Instruction **24**, 467–563 (2006). **(SD)** Students' conceptual-change processes around thermal equilibrium were mapped over time. Students maintained multiple contradictory ideas for extended periods of time, often as a result of experientially supported ideas that are different from school-supported ideas. (A)

## E. Chemical bonding and chemical energy

These papers deal with the energy associated with the formation and breaking of bonds in chemical reactions. Not surprisingly, most of them are located in the chemistry education literature, but we hope this Resource Letter will make this work known to the physics, engineering, and biology education communities, to facilitate incorporating this content into interdisciplinary thermodynamics education. "Chemical energy" is a black box in most introductory physics curricula (as well as in biology to some extent), and this chemistry education literature (and some biology) closely examines students' ideas about energy in chemical reactions. A central conceptual difficulty for many students is the idea that energy is "stored in bonds" (or that breaking bonds releases energy); another key issue is the relationship between the energy associated with chemical bonds and the energy of the rest of the system.

## 1. Biology

80. "ATP: A coherent view for school advanced level studies in biology," C. Gayford, J. of Biol. Educ. **20**, 27–32 (1986). **(SD)** Biology texts use the misleading language of "high-energy bonds" in explaining ATP hydrolysis. This paper explains the energetics of ATP hydrolysis and shows that this topic causes difficulties for biology students, especially those who have not taken chemistry. (E)

81. "Textbook errors & misconceptions in biology: Cell energetics," R.D. Storey, The American Biology Teacher **54**, 161–166 (1992). This article addresses problematic explanations of bioenergetics in textbooks, including "high-energy bonds" in ATP, and ATP as an energy storage compound. (E)

## 2. Chemistry

82. "Chemical energy: a learning package," I. Cohen and R. Ben-Zvi, J. Chem. Educ. **59**, 656–658 (1982). Gains in understanding chemical energy were achieved with a learning package that focuses on defining the system and the surroundings. (E)

83. "A new road to reactions. Part 3. Teaching the heat effect of reactions," W. de Vos and A.H. Verdonk, J. Chem. Educ. **63**, 972–974 (1986). Discusses laboratory experiments involving chemical reactions where a spontaneous change in temperature can be observed. (E)

84. "There is no energy in food and fuels - but they do have fuel value," K.A. Ross, School Science Review **75**, 39–47 (1993). Energy is not "in fuel," but associated with the fuel-oxygen system. Instructional implications include distinguishing between matter and energy. (E)

85. "Students' understandings of chemical bonds and the energetics of chemical reactions," H.K. Boo, J. Res. Sci. Teach. **35**, 569–581 (1998). **(SD)** Students saw a chemical bond as a physical entity; thus energy is required to form a bond, and breaking a bond releases energy. (E)

86. "Undergraduate students' understanding of enthalpy change," E.M. Carson and J.R. Watson, University Chemistry Education **3**, 46–51 (1999). **(SD)** Looks at first-year undergraduates' ideas about enthalpy and its role in thermodynamic theory. (I)

87. "Students' reasoning about basic chemical thermodynamics and chemical bonding: what changes occur during a context-based post-16 chemistry course?" V. Barker and R. Millar, Int. J. Sci. Educ.

conservation of energy, applied to chemical systems. Student data showed that this model forced students to grapple with conceptual issues. (E)

94. "Student conceptions about energy transformations: progression from general chemistry to biochemistry," A.J. Wolfson, S.L. Rowland, G.A. Lawrie, and A.H. Wright, Chem. Educ. Res. Pract. **15**, 168–183 (2014). **(SD)** Chemistry and biology students were interviewed and surveyed to determine the concepts around energy that they bring into biochemistry. Learning progressions are identified for concepts including the dependence of free energy changes on reaction conditions, the interpretation of energy diagrams, and the difference between standard and biological conditions. (E)

See also: Ref. 44

### 3. Physics

95. "Chemical energy in an introductory physics course for life science students," B.W. Dreyfus, J. Gouvea, B.D. Geller, V. Sawtelle, C. Turpen, and E.F. Redish, Am. J. Phys. **82**, 403–411 (2014). **(SD)** A curricular thread in an introductory physics course that uses chemical energy to build interdisciplinary connections among physics, chemistry, and biology. (E)

### 4. Multidisciplinary

96. "Structural characteristics of university engineering students' conceptions of energy," X. Liu, J. Ebenezer, and D.M. Fraser, J. Res. Sci. Teach. **39**, 423–441 (2002). **(SD)** As a follow-up to Ref. 90, concept maps for energy based on students' written paragraphs indicated difficulty in applying general principles such as conservation of energy to specific events in solution processes. Students demonstrated understanding of energy transformation and conservation, but not transport or degradation. (E)

## F. First law of thermodynamics

In biology and chemistry texts, the first law of thermodynamics is often synonymous with the law of conservation of energy. In physics and engineering, while these laws are of course equivalent, the first law refers to a particular formulation of this principle, stating that the change in the energy of a system is equal to the net energy that enters and leaves the system through heat and work processes. The latter definition of the first law defines this section of the Resource Letter, which includes papers dealing with the relationships between heat, work, and energy.

We note that the chemistry papers in this section contain little content that is specific to chemistry. In most cases, while they come from chemistry settings, they would not be out of place in physics settings. Biology is absent from this section entirely, suggesting that these concepts have not been emphasized in biology education research. In both physics and chemistry, the student data comes from contexts that are either abstract (removed from a specific physical situation) or from the standard introductory physics and engineering world of pistons and heat engines. In all the disciplines, we see an absence of research that engages with chemical reactions and biological processes.

### 1. Chemistry

97. "General definitions of work and heat in thermodynamic processes," E.A. Gislason and N.C. Craig, J. Chem. Educ. **64**, 660–668 (1987). Thermodynamic work and heat are defined operationally, based on the energy change in the surroundings. (E)

98. "Thermodynamics should be built on energy-not on heat and work," G.M. Barrow, J. Chem. Educ. **65**, 122–125 (1988). Heat and work are problematic constructs, and should be replaced with the change in energy of the thermal surroundings and of the mechanical surroundings. This approach leads to the first law and to the conservation of entropy for reversible processes. (E)

99. "'Work' and 'Heat': on a road towards thermodynamics," P.H. van Roon, H.F. van Sprang, and A.H. Verdonk, Int. J. Sci. Educ. **16**, 131–144 (1994). **(SD)** Students use "work" as a mechanical concept and "heat" as an energy concept. These can be converted into one another in the "thermochemical context," which is a precursor to the thermodynamic context. (E)

100. "'I believe I will go out of this class actually knowing something': Cooperative learning activities in physical chemistry," M.H. Towns and E.R. Grant, J. Res. Sci. Teach. **34**, 819–835 (1997). **(SD)** While this study focuses on the cooperative learning environment in a course for chemistry graduate students, it also engages with conceptual issues around $PV$ diagrams and cyclic processes. (E)

For an Engineering article see Ref. 9

## *2. Physics*

the mechanical coupling between them. In nonstatic expansions, there is acceleration, so work is difficult to calculate mechanically. (E)

107. "Developing the energy concepts in introductory physics," A.B. Arons, Phys. Teach. **27**, 506–517 (1989). The impulse-momentum and work-kinetic energy theorems are only valid for point masses; for deformable objects, the first law is needed. The first law (conservation of energy) is separated out from the center-of-mass equation, and a number of examples are considered. (E)

108. "General, restricted and misleading forms of the First law of thermodynamics," G.S. Moore, Phys. Educ. **28**, 228–237 (1993). Total energy ($E$) is not necessarily equal to internal energy ($U$), so some versions of the first law are misleading. Sample problems are provided to incorporate nonequilibrium thermodynamics qualitatively into introductory courses. (E)

109. "Development of energy concepts in introductory physics courses," A.B. Arons, Am. J. Phys. **67**, 1063–1067 (1999). Similar ideas to Ref. 107 emphasizing the difficulties that can be resolved with the use of the first law. (E)

110. "Difficulties in learning thermodynamic concepts: Are they linked to the historical development of this field?" M.I. Cotignola, C. Bordogna, G. Punte, and O.M. Cappannini, Science & Education **11**, 279–291 (2002). Student difficulties about heat and internal energy are analyzed in terms of the historical development of those concepts. Textbooks also display confusion between heat and internal energy. (E)

111. "Student understanding of the first law of thermodynamics: Relating work to the adiabatic compression of an ideal gas," M.E. Loverude, C.H. Kautz, and P.R.L. Heron, Am. J. Phys. **70**, 137–148 (2002). **(SD)** In a task involving the adiabatic compression of an ideal gas, students failed to use the first law, instead misapplying ideal gas concepts. Specific difficulties are enumerated, including confusion among work, heat, temperature, and internal energy. (E)

112. "Investigation of students' reasoning regarding heat, work, and the first law of thermodynamics in an introductory calculus-based general physics course," D.E. Meltzer, Am. J. Phys. **72**, 1432–1446 (2004). **(SD)** Students correctly used the state function concept in the context of energy, and incorrectly applied it to heat and work, which are not state functions. Students expressed the belief that the net work or net heat in a cyclic process is zero. (E)

## G. Entropy and the second law

The literature on entropy and the second law ranges from the practical, as in the development of tutorials aimed at addressing student difficulties with heat engines and the Carnot cycle, to the linguistic, as in the role played by language and conceptual metaphors in the construction of ideas about entropy. In the case of the latter, authors examine various metaphors by exploring their use in toy problems and simplified examples. In particular, considerable attention has been paid to the idea of treating entropy in terms of the "spreading" of energy, and to how this approach differs from those that describe entropy in terms of "disorder."

The physics literature focuses largely on the relationship of entropy to ideas surrounding either reversibility or energy, whereas the chemistry literature is primarily interested in the way in which an understanding of entropy and the second law contributes to a more thorough understanding of chemical equilibria. There is very little biology education literature focusing on the second law, either at the micro or macro scale.

We have chosen not to include here the significant number of papers addressing student difficulties at the advanced undergraduate or graduate level, particularly those detailing specific issues relating to statistical mechanics that would not likely be touched on in an introductory undergraduate curriculum.

### 1. Biology

119. "Molecular thermodynamics for cell biology as taught with boxes," L.S. Mayorga, M.J. Lopez, and W.M. Becker, CBE Life Science Education, **11**(1), 31–38 (2012). Proposes a model consisting of boxes with different shapes that contain small balls that are in constant motion due to a stream of air blowing from below. With such boxes, the basic concepts of entropy, enthalpy, and free energy can be taught while reinforcing a molecular understanding of these concepts. (E)

See also: Ref. 1

## *2. Chemistry*

120. "Misconceptions in school thermodynamics," A.H. Johnstone, J.J. Macdonald, and G. Webb, Phys. Educ. **12**, 248–251 (1977). **(SD)** Discusses conceptual difficulties encountered by students in reasoning about chemical equilibrium for the Scottish Certificate of Education. Particular attention is given to student difficulties surrounding Gibbs free energy. (I)

121. "Heat-fall and entropy," J.P. Lowe, J. Chem. Educ. **59**, 353 (1982). Describes how one can treat entropy in terms of "heat fall," where the fraction of heat energy that can be converted to work is the same as the fraction of the distance to absolute zero that the (remaining) heat falls. (E)

122. "Student misconceptions in thermodynamics," M.F. Granville, J. Chem. Educ. **62**, 847–848 (1985). **(SD)** Catalogs the most common thermodynamic misconceptions encountered by chemistry students having taken a semester of junior-level chemical thermodynamics. (I)

123. "Entropy analyses of four familiar processes," N.C. Craig, J. Chem. Educ. **65**, 760–764 (1988). Description of four processes is given in terms of entropy: a chemical reaction, a heat engine, the dissolution of a solid, and osmosis. (E)

124. "Entropy and the 2nd principle of thermodynamics - fourth year undergraduates' ideas," M.G.T.C. Ribeiro, Research in Assessment – Royal Society of Chemistry **IX**, 23–36 (1992). **(SD)** Looks at Portuguese undergraduates' reasoning about entropy and the second law of thermodynamics during semistructured interviews in the presence of live demonstrations. (I)

125. "A model of thermal equilibrium: A tool for the introduction of thermodynamics," R. Ben-Zvi, J. Silberstein, and R. Mamick, J. Chem. Educ. **71**, 31–34 (1993). **(SD)** Develops a new way of introducing thermodynamics that yields greater successes in student understanding of the distinction between heat and temperature. The new approach focuses on counting the ways in which systems can hold energy. (E)

For an article from Engineering see Ref. 9

## 3. Physics

133. "The second law of thermodynamics: a teaching problem and an opportunity," J. Ogborn, School Science Review **57**, 654–672 (1976). One of the earliest papers to address the difficulties inherent in teaching the second law of thermodynamics. General suggestions for how to introduce the topic, and commonly encountered pitfalls are discussed. (E)

134. "Teaching the approach to thermodynamic equilibrium: Some pictures that help," R. Baierlein, Am. J. Phys. **46**, 1042–1045 (1978). Provides a pictorial method for teaching students about the approach to equilibrium. The goal is to provide a more intuitive framework by drawing on familiar analogous ideas. (E)

135. "The role of the second law of thermodynamics in energy education," U. Haber-Schaim, Physics Teacher **21**, 17–20 (1983). Argues that energy education, as part of science education generally, should emphasize that reversing processes taking place spontaneously requires a fuel-consuming device. It calls for a greater role for thermodynamics in courses after studying energy conservation. (E)

136. "The second law of thermodynamics in a historical setting," J. Strnad, Phys. Educ. **19**, 94–100 (1984). Provides a historical overview of the development of thermodynamics generally, and the second law in particular, and argues that one should teach the material with this history in mind. Suggests that different presentations of the law are appropriate for different audiences. (E)

137. "Entropy in the teaching of introductory thermodynamics," H.U. Fuchs, Am. J. Phys. **55**, 215–219 (1987). By building on prescientific ideas about thermal phenomena, including the idea that heat is a substance contained in physical systems, the paper argues that it is possible to teach entropy effectively in introductory courses, and suggests how one might do so. (E)

138. "Students' understanding of basic ideas of the second law of thermodynamics," R. Duit and S. Kesidou, Research in Science Education **18**, 186–195 (1988). **(SD)** Explores how high school students take up qualitative ideas related to the second law. In particular, energy degradation and distribution, irreversibility and asymmetry, and the destructive aspects of the second law are considered. (I)

Explores student difficulties in applying the second law to cyclic systems like heat engines and refrigerators, and suggests tutorials that can help improve understanding. (E)

147. "Comment on 'Development and assessment of research-based tutorials on heat engines and the second law of thermodynamics,' by Matthew Cochran and Paula Heron," M. Bucher, Am. J. Phys. **75**, 377–378 (2007). **(SD)** This comment on Ref. 146 argues that the findings in that reference stem from a focus on the "pipeline" diagram that portrays heat as a fluid and emphasizes conservation of energy. (E)

148. "What is a reversible process?" M. Samiullah, Am. J. Phys. **75**, 608–609 (2007). An operational definition of reversible processes in terms of entropy is presented. The constancy of entropy, which defines reversible processes, also distinguishes such processes from those that are quasi-static. (E)

149. "Entropy, its language, and interpretation," H.S. Leff, Found. Phys. **37**, 1744–1766 (2007). Argues for use of the "spreading" metaphor (both spatially and temporally) for entropy, over other commonly used metaphors like "disorder" and "information." The paper provides examples illustrating why this treatment might be preferable. (E)

150. "Student ideas regarding entropy and the second law of thermodynamics in an introductory physics course," W. Christensen, D.E. Meltzer, and C.A. Ogilvie, Am. J. Phys. **77**, 907–917 (2009). **(SD)** Reports on student thinking about entropy in an introductory physics course, and shows that the conception of entropy as a conserved quantity is widespread. (I)

151. "Addressing student difficulties with concepts related to entropy, heat engines, and the Carnot cycle," T.I. Smith, W.M. Christensen, and J.R. Thompson, *AIP Conf. Proc.* **1179**, 277–281 (2009). **(SD)** Assessment data suggest that student understanding of heat engines and the Carnot cycle is improved by implementation of a guided-inquiry tutorial related to these topics. (I)

152. "Introducing thermodynamics through energy and entropy," R. de Abreu and V. Guerra, Am. J. Phys. **80**, 627–637 (2012). Proposes introducing thermodynamics with the concept of internal energy of deformable bodies. By introducing entropy before the notions of temperature and heat, the approach is meant to avoid some of the major conceptual difficulties with the traditional presentation. (E)

## 4. Multidisciplinary

160. "Textbook Forum. Thermodynamics of 'mixing' of ideal Gases: A persistent pitfall," E.F. Meyer, J. Chem. Educ. **64**, 676–677 (1987). Discusses the commonly held belief that mixing ideal gases causes an increase in entropy. (E)

161. "Entropy and the shelf model: A quantum physical approach to a physical property," A.H. Jungermann, J. Chem. Educ. **83**, 1686–1694 (2006). The concept of atomic entropy is introduced so that entropy values of substances with different stoichiometry may be compared much more rationally than on the basis of the values of molar entropy. (E)

162. "Energy diagrams for enzyme-catalyzed reactions: Concepts and misconcepts," J.C. Aledo, C. Lobo, and A.E. del Valle, Biochemistry and Molecular Biology Education **31**, 234–236 (2003). **(SD)** Suggests that for an enzyme-catalyzed reaction, textbooks should emphasize that under conditions where the overall reaction is spontaneous, each elementary step must exhibit a negative free-energy change. This must then be properly reflected in the progression profile of reaction diagrams. (E)

## 5. Non-discipline-specific

163. "Energy and fuel: The meaning of 'the go of things,'" J. Ogborn, School Science Review **68**, 30–35 (1986). Takes up the common belief that the possession of energy drives, gives potential for, or accounts for change. Free energy or entropy then represents the *possibility* of change. (I)

164. "Matter scatter and energy anarchy: The second law of thermodynamics is simply common experience," K.A. Ross, School Science Review **69**, 438–445 (1988). Takes the view that the second law of thermodynamics is uniquely rooted in everyday experience, and as such should be taught before the first law. (I)

165. "Scientific mental representations of thermodynamics," C. Tarsitani and M. Vicentini, Science & Education **5**, 51–68 (1996). Explains how the attitudes toward thermodynamics conveyed in commonly used textbooks underlie the relationship between the macroscopic and the microscopic approach on one side and between the "state" or "process" approach on the other. (E)

166. "Shuffled cards, messy decks, and disorderly dorm rooms—examples of entropy increase? Nonsense!" F.L. Lambert, J. Chem. Educ. **76**, 1385–1387 (1999). Explains that no permanent entropy change occurs in a macroscopic object after it has been transported from one location to another or when a group of them is scattered randomly. (E)

**H. Probability/Statistics and the second law**

Some literature relating to student understanding of the second law of thermodynamics focuses not on specific curricular topics, but on student competency with ideas of probability and statistics that are essential components of any comprehensive treatment of entropy. At least one paper in biology education focuses on the role that student understanding of randomness plays in the development of a coherent conceptual model of the second law. The absence of a more extensive biology education literature on the subject is striking given the central role that randomness plays in biological systems.

*1. Biology*

173. "Understanding randomness and its impact on student learning: Lessons learned from building the Biology Concept Inventory (BCI)," K. Garvin-Doxas and M.W. Klymkowsky, CBE Life Science Education **7**(2), 227–233 (2008). **(SD)** Describes how a wide class of student difficulties in molecular and evolutionary biology may be based on deep-seated misconceptions about random processes. For example, most students believe that diffusion takes place only when there is a concentration gradient. (E)

*2. Chemistry*

174. "An integrated, statistical molecular approach to the physical chemistry curriculum," S.F. Cartier, J. Chem. Educ. **86**, 1397–1402 (2009). Argues against a compartmentalized approach to physical chemistry, in which thermodynamic and molecular concepts are treated separately. An integrated curriculum requires consideration of entropy in the microscopic realm. (E)

*3. Physics*

175. "Student estimates of probability and uncertainty in advanced laboratory and statistical physics courses," D.B. Mountcastle, B. Bucy, and J.R. Thompson, *AIP Conf. Proc.* **951**, 152–155 (2007). **(SD)** Looks at statistical physics students' reasoning about the relative uncertainties of binary outcomes, showing that students did not reliably recognize that this uncertainty goes down as the number of measurements increased. (E)

176. "Student understanding of basic probability concepts in an upper-division thermal physics course," M.E. Loverude, *AIP Conf. Proc.* **1179**, 189–192 (2009). **(SD)** Diagnostic questions probed student understanding of probability concepts, showing that students struggled in distinguishing microstates from macrostates and in using mathematics to describe the multiplicity of a system. (I)

177. "Investigating student understanding for a statistical analysis of two thermally interacting solids," M.E. Loverude, *AIP Conf. Proc.* **1289**, 213–216 (2010). **(SD)** Describes a series of tutorials in which undergraduate students apply statistical methods to examine the behavior of two interacting Einstein solids. Strengths and weaknesses of student reasoning, both qualitative and quantitative, are explored. (I)

## I. Free energy and the second law

Much of the chemistry education literature on entropy surrounds the role that it plays in determining the spontaneity of chemical processes. In particular, the connection between entropy and Gibbs free energy is discussed extensively, as is the degree to which an understanding of entropy aids in an understanding of enthalpy. The relationship among these three constructs – entropy, enthalpy, and Gibbs free energy – forms the cornerstone of student understanding of the second law as it relates to chemistry, and of an understanding of chemical equilibria. Physics education literature relating to these ideas is notably absent.

### 1. Chemistry

178. "Reaction and spontaneity: the influence of meaning from everyday language on fourth year undergraduates' interpretation of some simple chemical phenomena," M.G.T.C. Ribeiro, D.J.V. Costa Pereira, and R. Maskill, Int. J. Sci. Educ. **12**, 391–401 (1990). **(SD)** Looks at undergraduate chemistry students' reasoning about the terms "reaction" and "spontaneous" in a variety of chemical contexts. Implications for university-level teaching are discussed. (I) thermodynamic entropy and stored internal energy in a solid are intimately related and that entropy can be usefully interpreted as an energy-spreading function, as described in Refs. 142 and 154–158. (I)

187. "Prospective chemistry teachers' conceptions of chemical thermodynamics and kinetics," M. Sözbilir, T. Pınarbaşı, and N. Canpolat, EURASIA Journal of Mathematics, Science and Technology Education **6**, 111–120 (2010). **(SD)** Looks at difficulties encountered by prospective chemistry teachers in Turkey in trying to distinguish reaction thermodynamics from reaction kinetics. (I)

See also: Ref. 91, Ref. 94, Ref. 98

### 2. Multidisciplinary

188. "Coupled reactions 'versus' connected reactions: coupling concepts with terms," J.C. Aledo, Biochemistry Molecular Biology Education **35**, 85–88 (2007). When considering coupled reactions, both students and textbook authors often make claims that clash with the second law of thermodynamics. This paper points out the most common flaws, analyzes the causes leading to these mistakes, and suggests a few rules to avoid them. (E)

## J. Osmosis, diffusion, and randomness

Although the biology education literature does very little to address osmosis and diffusion in a mechanistic way, the literature does describe student understanding of these ideas from a phenomenological perspective. Many papers explore the success or lack thereof of various interventions and pedagogical instruments in improving student mastery of core concepts surrounding diffusion and osmosis in detail. The authors of these papers may or may not themselves view these topics as falling under the "thermodynamics" umbrella, but we include them owing to their underpinning in statistical physics. The physics and chemistry education literature on diffusion and osmosis is notably limited.

# 1. *Biology*

189. "Problem solvers' conceptions about osmosis," J.T. Zuckerman, American Biology Teacher **56**, 22–25 (1994). **(SD)** Discusses the scheme and findings of a study designed to identify the conceptual knowledge used by high school students to solve a problem related to osmosis. Tips are provided to teachers to aid students. (E)

190. "Students' misconceptions about diffusion: How can they be enhanced," E.A. Marek, *et al.*, American Biology Teacher **56**, 74–77 (1994). **(SD)** Describes a study designed to seek the causes for student misconceptions surrounding diffusion, and to find ways to eliminate them. (E)

191. "Dealing honestly with diffusion," S. Vogel, American Biology Teacher **56**, 405–407 (1994). Identifies common misconceptions regarding diffusion that exist among many biology teachers as well as students. Offers suggestions and demonstrations to use in the classroom to help students gain a more accurate understanding of diffusive processes. (E)

192. "Accurate and inaccurate conceptions about osmosis that accompanied meaningful problem solving," J.T. Zuckerman, School Science and Mathematics **94**, 226–234 (1994). **(SD)** Discusses some accurate and inaccurate conceptions about osmosis that were identified in interviews of 16 outstanding science students. (E)

193. "Secondary & college biology students' misconceptions about diffusion & osmosis," A.L. Odom, American Biology Teacher **57**, 409–415 (1995). **(SD)** Describes tests on diffusion and osmosis that were given to 116 secondary biology students, 123 university-level non-biology majors, and 117 university-level biology majors. Students continued to have misconceptions about these ideas, even after instruction. (I)

194. "Development and application of a two-tier diagnostic test measuring college biology students' understanding of diffusion and osmosis after a course of instruction," A.L. Odom and L.H. Barrow, J. Res. Sci. Teach. **32**, 45–61 (1995). **(SD)** Describes a diagnostic test for measuring college biology students' understanding of diffusion. (I)

195. "Reduce confusion about diffusion," M.R. Hebrank, American Biology Teacher **59**, 160–163 (1997). Describes activities that allow students to explore diffusion by appealing to their kinesthetic senses. Also presents a computer simulation of diffusion. (E)

202. "Teaching diffusion with a coin," H. Haddad H and M.V.C. Baldo, Advances in Physiology Education **34**, 156–157 (2010). Describes an inexpensive and simple way for students to experience the probabilistic and random motion of diffusing particles. (E)

203. "Osmosis and diffusion conceptual assessment," K.M. Fisher, K.S. Williams, and J.E. Lineback, CBE - Life Sciences Education **10**, 418–429 (2011). To monitor comprehension of osmotic and diffusive processes among students at a large public university, the authors developed and validated an 18-item Osmosis and Diffusion Conceptual Assessment (ODCA). This assessment includes two-tiered items, some adopted or modified from the previously published Diffusion and Osmosis Diagnostic Test (DODT) and some newly developed items. (I)

## 2. Physics

204. "Five popular misconceptions about osmosis," E.M. Kramer and D.R. Myers, Am. J. Phys. **80**, 694–699 (2012). While more advanced than is likely appropriate for many introductory students, this paper develops ideas about osmosis from first principles of statistical mechanics. It addresses common misconceptions that students have about osmosis, and why these are likely to arise out of the formalism. (E)

## 3. Multidisciplinary

205. "An interactive computer model for improved student understanding of random particle motion and osmosis," J. Kottonau, J. Chem. Educ. **88**, 772–775 (2011). Simulations are developed to help students understand that the membrane-crossing probability of water molecules depends solely on their concentrations on both sides of the membrane. (E)

## 4. Non-discipline-specific

206. "Demonstrating diffusion: Why the confusion?" D.L. Panizzon, Australian Science Teachers' Journal **44**, 37–39 (1998). Examines how the process of diffusion may be confused with convection. (E)

207. "Using a cognitive structural model to provide new insights into students' understandings of diffusion," D. Panizzon, Int. J. Sci. Educ. **25**, 1427–1450 (2003). **(SD)** Describes a pathway of conceptual understanding of diffusion from simple intuitive ideas about movement to highly abstract views in which students explained the random motion of molecules in terms of kinetic theory. (I)

## IV. CONCLUSIONS

At present, education research in physics, chemistry, and biology has come to focus on understanding how to better teach issues in thermodynamics. Similarly, the corresponding fields of discipline-based education research have begun to explore an understanding of how introductory-level undergraduate students understand the concepts central to thermodynamics. However, we see little evidence that the research is drawing across disciplinary boundaries to make progress in this arena. Our aim in this Resource Letter has been to benefit instructors and researchers in these three domains, and to draw attention to places where coordination between and among disciplines would be fruitful.

Additionally, a rising interest in Introductory Physics for the Life Sciences (IPLS) courses requires an attention to thermodynamics and student understanding of thermodynamics. We intend the collection presented here as an aid to instructors seeking to increase interdisciplinary connections in developing IPLS courses. We encourage all of these audiences to be more attentive to cross-disciplinary avenues of research in the search for understanding and improving students' conceptual ideas in thermodynamics. The greatest progress in our goals will be made by coordinating resources, focusing on how students understand ideas within thermodynamics, and talking across disciplinary boundaries.

## ACKNOWLEDGMENTS

The authors thank Edward F. Redish for his invaluable assistance in preparing this Resource Letter. We also thank Chandra Turpen and Julia Gouvea for their support in articulating methods for the literature search. We are grateful for the contributions of Chris Bauer, Melanie Cooper, Catherine Crouch, and Mike Klymkowsky for aid in providing insight and research expertise into disciplinary perspectives from biology, chemistry, and physics education. We also thank the University of Maryland Physics Education Research Group and Biology Education Research Group for their contributions to this work. This work is supported by an NSF Graduate Research Fellowship (DGE 07-50616), NSF-TUES DUE 11-22818, and the Howard Hughes Medical Institute NEXUS grant. Any opinions, findings, conclusions, or recommendations expressed in this publication are those of the authors and do not necessarily reflect the views of the National Science Foundation or the Howard Hughes Medical Institute.